\documentclass[preprint,11pt,numbers,sort&compress]{elsarticle}
\pdfoutput=1
\PassOptionsToPackage{table}{xcolor}
\usepackage{geometry}
\usepackage[T1]{fontenc}
\usepackage[utf8]{inputenc}
\usepackage{caption}
\usepackage{lmodern}
\usepackage{mathpazo}
\usepackage[varqu,varl,var0,scaled=0.97]{inconsolata}

\usepackage[colorlinks = true]{hyperref}
\usepackage{amsmath,amsfonts,amssymb}
\usepackage{url}
\makeatletter
\g@addto@macro{\UrlBreaks}{\UrlOrds}
\makeatother
\usepackage{mathtools}
\usepackage{xcolor}
\usepackage{slashed}
\usepackage{xspace}
\usepackage[absolute]{textpos} 
\usepackage{longtable}

\journal{Computer Physics Communications}

\newcommand*{\cf}{cf.\ }
\newcommand*{\eg}{e.\,g.\ }
\newcommand*{\ie}{i.\,e.\ }
\newcommand*{\Eq}{Eq.\,}

\definecolor{identifiercolor}{rgb}{.4,.6,.56}
\definecolor{stringcolor}{gray}{0.5}
\definecolor{inactivecolor}{rgb}{0.15,0.15,0.5}
\usepackage{listings}
\begin{document}
\begin{frontmatter}

\title{FeynCalc 10.2 and FeynHelpers 2: Multiloop calculations streamlined}

\author[a]{Vladyslav Shtabovenko\texorpdfstring{\corref{author1}}{}}

\cortext[author1] {\textit{E-mail address:} shtabovenko@physik.uni-siegen.de}

\address[a]{Theoretische Physik 1, Center for Particle Physics Siegen,  
Universität Siegen, \\ Walter-Flex-Str. 3, 57068 Siegen, Germany}

\begin{textblock*}{100ex}(0.8\textwidth,5ex)
{P3H-25-112, SI-HEP-2025-31}
\end{textblock*}

\begin{abstract}

We present new versions of the  \textsc{Mathematica} package \textsc{FeynCalc} and the 
\textsc{FeynHelpers} add-on that represent an important contribution to the collection of public codes for semi-automatic evaluation of multiloop Feynman diagrams. \textsc{FeynHelpers} enables 
interfacing between \textsc{FeynCalc} and selected programs routinely used in 
multiloop calculations such as \textsc{QGRAF}, \textsc{FIRE}, \textsc{KIRA}, \textsc{FIESTA} or \textsc{pySecDec}. In addition to that
\textsc{FeynCalc} now supports parallelizing the execution of selected routines on multicore
CPUs and can handle tensor integrals with zero Gram determinants where the traditional tensor
reduction breaks down.
\end{abstract}

\end{frontmatter}

{\bf PROGRAM SUMMARY/NEW VERSION PROGRAM SUMMARY}

\begin{small}
    \noindent
    {\em Program Title:} FeynCalc and FeynHelpers                                         \\
    {\em CPC Library link to program files:} (to be added by Technical Editor) \\
    {\em Developer's repository link:} \url{https://github.com/FeynCalc/feyncalc}, \url{https://github.com/FeynCalc/feynhelpers} \\
    {\em Code Ocean capsule:} (to be added by Technical Editor)\\
    {\em Licensing provisions:} GPLv3 \\
    {\em Programming language:} Wolfram Language                                  \\
    {\em Supplementary material:} Manual, example notebooks                        \\
    {\em Journal reference of previous version:} Comput. Phys. Commun., \textbf{306}, 109357, 2024; 
        Comput. Phys. Commun., 218 (2017) 48-65 \\
    {\em Does the new version supersede the previous version?:} Yes  \\
    {\em Reasons for the new version:} More interfaces to useful programs for multiloop calculations as well as improvements of the existing algorithms and support for parallel evaluation \\
    {\em Summary of revisions:} Complete multiloop calculations can be now carried out from a \textsc{Mathematica} notebook using 
    \textsc{FeynCalc}.\\
    {\em Nature of problem:} Analytic calculations of higher-order corrections to particle physics processes using Feynman diagrammatic expansion.\\
    {\em Solution method:} Part of the required algorithms and algebraic identities are implemented in Wolfram \textsc{Mathematica}. Functionality offered through external programs can be accessed using a collection of interfaces.\\
    {\em Restrictions:} Depending on the complexity of the problem, the number of terms might exceed the capacities of  \textsc{Mathematica}.
    
\end{small}

\section{Introduction}

While the much anticipated new physics is yet to be discovered, high-energy theory community actively continues its work of meticulously studying theoretical implications of the Standard Model and cross checking them against experimental measurements. Here multiloop calculations remain among the key tools aimed at increasing the precision of theoretical predictions for numerous observables.

Over the past decades multiloop community has achieved remarkable progress in developing new algorithms (\cf \eg~\cite{Smirnov:2006ry,Weinzierl:2022eaz,Smirnov:2025dfy}) and creating powerful software tools aimed at doing calculations that seemed to be unfeasible or only doable only by a handful of experts in the past. Moreover, nowadays many modern laptops offer computational capabilities that were once the domain of high-end workstations. As a result, many calculations (\eg at two or three loops with one or two scales) that previously demanded substantial hardware resources are today well within reach even for researchers without regular access to large computing clusters.

Despite the rapid progress in computational techniques, many of these developments remain unfamiliar to a large portion of the theory community. Often, researchers might have heard that particular codes suitable for tackling the given task exist, but the effort required to install them, learn their usage, and adapt them to a specific project may appear too complicated and time-consuming. This is why it becomes increasingly important to develop user-friendly and easy-to-use tools that lower the entry barrier to the field of multiloop calculations. The goal is empowering researchers, especially early-career ones, to perform computations beyond one loop without spending excessive time on coding and interfacing different programs with each other.

In this context \textsc{Mathematica} package \textsc{FeynCalc}~\cite{Mertig:1990an,Shtabovenko:2016sxi,Shtabovenko:2020gxv,Shtabovenko:2021hjx,Shtabovenko:2021hjx,Shtabovenko:2023idz}
has earned itself a reputation of an accessible, practical and versatile tool suitable for both getting familiar with Feynman diagram calculations and doing high-profile research. Towards the end of 2023, following an extensive testing phase~\cite{Shtabovenko:2021hjx} the program was finally extended to support multiloop calculations~\cite{Shtabovenko:2023idz}. However, \textsc{FeynCalc} 10 still did not cover all steps typically encountered when evaluating unrenormalized Feynman diagrams beyond one loop. In particular, the connection to tools for Integration-By-Parts (IBP)~\cite{Chetyrkin:1981qh,Tkachov:1981wb} reduction and numerical  evaluation of master integrals was lacking, while the diagram generation remained limited to \textsc{FeynArts}~\cite{Hahn:1998yk,Hahn:2000kx}.

In the present work we present a collection of interfaces between \textsc{FeynCalc} and various external software packages that close this gap. When used together with the \textsc{FeynHelpers} 2 add-on, \textsc{FeynCalc} now allows the user to reduce the given amplitude to a set of master integrals and to evaluate those integrals numerically in a straightforward and convenient manner. In comparison to the previous version of \textsc{FeynHelpers} that only offered interfaces to \textsc{Package-X}~\cite{Patel:2015tea,Patel:2016fam} and the \textsc{Mathematica} version of \textsc{FIRE}~\cite{Smirnov:2019qkx,Smirnov:2023yhb,Smirnov:2025prc}, the list of supported tools have been significantly extended.

First, it is now possible to generate Feynman diagrams using \textsc{QGRAF}~\cite{Nogueira:1991ex} and directly import the expressions for the amplitudes into \textsc{FeynCalc}. Visualization of the diagrams can be done with \textsc{GraphViz}~\cite{url:Graphviz} or \textsc{TikZ-Feynman}~\cite{Ellis:2016jkw}. As far as the IBP-reduction is concerned, loop integrals and topologies obtained from the amplitude in the \textsc{FeynCalc} format (\ie as \texttt{GLI} and \texttt{FCTopology} containers) can be directly converted into run cards for \textsc{FIRE} + \textsc{LiteRed}~\cite{Lee:2012cn,Lee:2013mka} or \textsc{KIRA}~\cite{Maierhofer:2017gsa,Klappert:2020nbg,Lange:2025fba}. Depending on the available computer resources, the reduction process can be carried out on a cluster or (for simple cases) done directly on the user's machine. The interface also handles the import of the so-obtained reduction tables into \textsc{Mathematica}. If needed, they can be directly inserted into the amplitude at hand. Numerical evaluation of the resulting master integrals is offered using \textsc{pySecDec}~\cite{Borowka:2017idc,Borowka:2018goh,Heinrich:2021dbf,Heinrich:2023til} or \textsc{FIESTA}~\cite{Smirnov:2015mct,Smirnov:2021rhf}. Just as in the case of the IBP-reduction, the automatically generated scripts can be either run locally or transferred to more powerful machines. Last but not least, \textsc{FeynHelpers} also supports the numerical evaluation of Passarino--Veltman functions using \textsc{LoopTools}~\cite{Hahn:1998yk} and can speed up solving of linear equation systems with \textsc{FERMAT}~\cite{Lewis:Fermat}. The latter is a crucial ingredient for calculating high-rank tensor reduction formulas using \textsc{FeynCalc}'s \texttt{Tdec} routine.

The evolution of the \textsc{FeynHelpers} interface also triggered some important improvements in \textsc{FeynCalc} that were overdue. One of them is a working solution for the notorious problem of zero Gram determinants in the tensor reduction of multiloop integrals. The other is the ability to use parallel evaluation for selected  \textsc{FeynCalc} routines.

The remainder of this paper is organized as follows. In Section~\ref{sec:multiloopauto} we outline our idea of multiloop automation
and give a brief overview of the current state of the arts. Installation of \textsc{FeynHelpers} is described in  
Section \ref{sec:install}, while Section \ref{sec:docu}  explains how to access the documentation of the package. A description of the new interfaces and the improvements in \textsc{FeynCalc} are  offered in Section \ref{sec:interfaces} and Section \ref{sec:fc} respectively. 
Last but not least,  in Section \ref{sec:examples} we showcase some example calculations that make use of the \textsc{FeynCalc} and \textsc{FeynHelpers} tandem. Finally, we summarize our results in Section \ref{sec:summary}.


\section{Multiloop automation: Trends and challenges} \label{sec:multiloopauto}

In order to avoid any misunderstandings or confusions let us first carefully explain what
is meant with \emph{multiloop automation} in the context of this paper. While all technical steps 
needed to obtain an observable (\eg decay rate or cross section) at one-loop accuracy \footnote{
We avoid using the notion of leading-order (LO) and next-to-leading-order (NLO) in this discussion
because there are processes that are LO at one-loop, \eg gluonic Higgs decay $h \to gg$ or 
neutral $B$-meson mixing $b \bar{s} \to s \bar{b}$} are sufficiently well understood to be automatized and implemented in public software frameworks (\cf \eg~\cite{Alwall:2014hca,Cullen:2011ac,GoSam:2014iqq,Bahr:2008pv,Bellm:2015jjp,Bevilacqua:2011xh,Nason:2004rx,Frixione:2007vw,Alioli:2010xd,Gleisberg:2008ta,Sherpa:2019gpd,Moretti:2001zz,Kilian:2007gr,Yuasa:1999rg,Fujimoto:2002sj,Hahn:1998yk} and many others), the situation at two loops and higher is more involved. While the community is slowly progressing towards general-purpose two-loop codes~\cite{Borowka:2016agc,Borowka:2016ehy,Pozzorini:2022ohr,Zoller:2022ewt,Canko:2023lvh,Abreu:2020xvt,Heinrich:2023til}, so far there are no packages similar to \eg \textsc{FormCalc}, \textsc{MadGraph} or \textsc{GoSam} for calculating arbitrary processes at two loops and above.

Nevertheless, some of the building blocks needed for a full multiloop calculation are very well suited for automation. Here we want to focus on the evaluation of a generic unrenormalized amplitude $i \mathcal{M}$ beyond one loop, where one can follow a recipe that we prefer to call a \emph{standard multiloop workflow}. The schematic steps behind this procedure are outlined below

\begin{enumerate}
\item Generation of Feynman diagrams using \textsc{QGRAF}, \textsc{FeynArts} or some private tool
\item Insertion of Feynman rules and possibly filtering of the diagrams according to some criteria
\item Evaluation of Dirac and color algebra, insertion of projectors or tensor reduction as well as additional
simplifications specific to the process at hand. In most cases this is done using \textsc{FORM}~\cite{Vermaseren:2000nd,Kuipers:2012rf,Ruijl:2017dtg}
but can be also handled by any sufficiently powerful symbolic manipulation system (\eg \textsc{Symbolica}~\cite{url:Symbolica})
\item Identification and minimization of loop integral families as well as rewriting of the explicit loop integrals into the notation where each integral is characterized only by its family identifier and propagator powers. To ensure that the set of propagators in each integral family forms a basis, it might be necessary to perform partial fraction decomposition or add auxiliary propagators. This step is usually handled using dedicated scripts or tools (\eg \textsc{feynson}~\cite{Maheria:2022dsq}, \textsc{Alibrary}~\cite{url:Alibrary} \textsc{exp}~\cite{Harlander:1998cmq,Seidensticker:1999bb}, \textsc{TopoID}~\cite{Hoff:2016pot}, \textsc{Tapir}~\cite{Gerlach:2022qnc}) that generate all the necessary replacement rules for \textsc{FORM}
\item Reduction of all loop integrals to master integrals using IBP techniques with the aid of 
\textsc{FIRE} + \textsc{LiteRed}, \textsc{LiteRed} alone, \textsc{KIRA}, \textsc{Blade}~\cite{Liu:2018dmc,Guan:2019bcx,Guan:2024byi}, \textsc{Reduze}~\cite{vonManteuffel:2012np} etc.
\item Manipulation and analysis of the reduction tables with the purpose of expanding them in $\varepsilon$, changing the master integral
basis~\cite{vonManteuffel:2014qoa,Smirnov:2020quc,Usovitsch:2020jrk}, deriving differential equations, uncovering one-to-one~\cite{Pak:2011xt} or linear~\cite{Davies:2025ghl,Davies:2024kvt,Davies:2022ram,Davies:2018qvx} relations
between masters etc. Such tasks can be conveniently done in \textsc{Mathematica}, even though one can use any other suitable computer algebra system. 
\item Insertion of optimized reduction rules into the amplitude, which is mostly done in \textsc{FORM} (or \textsc{Symbolica}) due to 
the large size and high complexity of the resulting intermediate expressions.
\end{enumerate}

Finally, after completing all the above steps one obtains the unrenormalized amplitude expressed in terms of a presumably minimal set
of master integrals. What happens after that (\eg evaluation of master integrals, renormalization, subtraction of infrared (IR) divergences etc.) is highly dependent on the given process and the goal of the calculation. In this paper we abstain from discussing the automation of those subsequent steps, because they usually turn out to be too specific and process-dependent to be reduced to a template. 

Contrary to that, steps outlined in the standard multiloop workflow happen to occur in the course of  almost any multiloop calculation. 
It is important to emphasize, however, that standard by no means implies trivial. Indeed, complications and unexpected bottlenecks can (and usually do) arise at almost every stage of the standard multiloop workflow. For example, even seemingly simple tasks such as algebraic simplifications of Dirac matrices can produce millions of intermediate terms and lead to a complete breakdown of the calculation unless handled carefully. IBP reduction of particularly complicated topologies may turn out to be unfeasible with the current technology and computation resources, requiring a complete redesign of the project. In addition to that, there are always numerous bottlenecks in different parts of the whole toolchain, which might have been irrelevant during the calculation of X but become a show stopper when trying to calculate Y. Nevertheless, as of now the standard multiloop workflow is the closest we can get to having publicly available codes that can automate calculations beyond one loop. 

In recent years we witnessed an increasing number of open-source tools (\cf Section 2 of ref.~\cite{Shtabovenko:2023idz} for a brief overview) implementing something along the lines of the standard multiloop workflow (possibly adding or omitting several steps) and thus making multiloop calculations more accessible to nonexperts: \textsc{tapir}, \textsc{Alibrary}, 
\textsc{FeAmGen.jl}~\cite{Wu:2023qbr}, \textsc{HepLib}~\cite{Feng:2021kha,Feng:2023hxy}, \textsc{MaRTIn}~\cite{Brod:2024zaz}, 
\textsc{ANATAR}~\cite{Duhr:2025oil}, \textsc{AmpRed}~\cite{Chen:2024xwt,Chen:2025paq}, \textsc{CalcLoop}~\cite{url:CalcLoop} and others. This is undoubtedly a very positive development for a field, where most practitioners still consider it perfectly normal to keep all computational codes private or only make them available to collaborators.

In this context, it is useful to highlight several common challenges arising when researchers decide to create a useful and sustainable framework and share it with the community.

A major difficulty, which is often underestimated, is that any framework built on top of multiple external programs requires continuous maintenance. Even minor syntax changes introduced in new versions of  \textsc{FORM}, \textsc{Mathematica}, \textsc{KIRA} and similar codes can create bugs or, in the worst case, render the entire framework unusable. Consequently, software for multiloop automation is not something one can publish once and then leave it to accumulate citations like a typical nonsoftware publication. To remain functional and useful for the community, such projects require at least one developer who actively uses the code in their own research and therefore can detect and fix issues, add new features, and, most importantly, ensure that everything continues to work as intended.

However, working conditions in academia differ substantially from those in the software industry, where people earn their money and reputation by writing code. Theorists who typically take on the task of providing community service in terms of useful codes are untenured doctoral or postdoctoral researchers. The time and effort invested into maintaining and improving software tools necessarily reduces the time available for producing research papers, which in turn may diminish their prospects of securing a permanent academic position and ultimately force them to leave academia for industry. After all, scientific software development is rarely regarded as ``doing physics'', regardless of how essential or widely used the resulting code may be in the work of other physicists. Conversely, researchers with long-term or permanent positions often have a long list of additional duties such as group management, teaching, grant applications, student supervision etc. These obligations come on top of their own research and leave little room for coding-related activities. As a consequence, many published codes end up in an unmaintained state and ultimately stop working.

Another common issue is limited flexibility. Multiloop frameworks often evolve from private codes initially written with a particular calculation in mind and later extended to be more versatile. However, the constraints of the initial design still remain and might cause problems, when the program is being used for a completely different type of calculations. For example, software originally developed for QCD processes with only one or two scales may be poorly optimized for electroweak calculations, which involve additional masses and subtleties such as the treatment of $\gamma^5$ in dimensional regularization. Likewise, calculations related to weak effective Hamiltonians~\cite{Buras:1998raa,Buras:2020xsm} introduce further complications due to four-fermion operators, evanescent operators~\cite{Herrlich:1994kh}, and related technicalities. The consequence is that any framework requires time to mature and become sufficiently refined to meet the day-to-day needs of its users. This process inevitably demands that the authors of the code invest effort into solving issues unrelated to their own research interests. Unfortunately, this is the only way to ensure that the framework can survive and acquire a sufficient user base rather than ending up as yet another abandoned tarball on a website.

Last but not least, authors often underestimate the importance of providing a sufficient number of clear, useful, and well-documented examples. The potential users of a multiloop framework are physicists with limited time and multiple projects running in parallel.
The quicker they can learn how to apply the program to their own calculations, the more likely they are to continue using it for future work. Fully worked-out, realistic examples are therefore essential, not only because they illustrate how to use the framework but also for the sake of having some sanity tests when fixing bugs or rewriting large portions of the code.  Unfortunately, preparing such examples for multiloop calculations quickly becomes a formidable and time-consuming task. At one loop, one can often rely on \textsc{Package-X} to evaluate master integrals analytically and easily compare with results from the literature. Yet already at two loops the situation becomes significantly more complicated. To begin with, there is nothing even remotely similar to a library of known analytic results for multiloop integrals. The Loopedia~\cite{Bogner:2017xhp} website might provide hints for papers containing the desired expressions, but does not spare the user from the tedious process of extracting formulas from \TeX\ sources,\footnote{Some authors kindly provide machine-readable data, but this is much less common than one might hope.} cross-checking them against numerical results\footnote{The recent release of Subtropica~\cite{Giroux:2026tgd} raises hopes that the availability of analytic results for master integrals may improve in future.}. Moreover, the number of master integrals appearing in two- or three-loop amplitudes can easily range from 
$\mathcal{O}(10)$ to $ \mathcal{O}(10^3)$ or more, while the basis used in the literature will almost certainly differ from the one automatically chosen by \textsc{FIRE} or \textsc{KIRA}. Apart from that, many authors provide only final renormalized results and refrain from listing intermediate expressions, such as unrenormalized amplitudes written explicitly in terms of master integrals.
All these complications make the task of working out explicit examples for an automation framework  both challenging and extremely time-consuming. In the end, one often spends considerable effort merely to reproduce results that were already computed and published long ago.

To summarize, we have discussed the current status of multiloop automation for what we refer to as the standard template and highlighted 
several common challenges encountered in such projects. In the following, we will now focus specifically on the \textsc{FeynCalc} and 
\textsc{FeynHelpers} framework that offers yet another approach to the task of automatizing multiloop calculations.

\section{Installation} \label{sec:install}

Being an add-on \textsc{FeynHelpers} obviously  requires \textsc{FeynCalc} to be installed first, which can be done running

\begin{lstlisting}[extendedchars=true,language=Mathematica]
    Import["https://raw.githubusercontent.com/FeynCalc/feyncalc/master/install.m"]
    InstallFeynCalc[]
\end{lstlisting}
\noindent on a fresh \textsc{Mathematica} kernel. All versions starting with 10 are supported, although we recommend using \textsc{Mathematica} 13 or 14 for the best user experience. Installing \textsc{FeynHelpers} amounts to evaluating

\begin{lstlisting}[extendedchars=true,language=Mathematica]
    Import["https://raw.githubusercontent.com/FeynCalc/feynhelpers/master/install.m"]
    InstallFeynHelpers[]
\end{lstlisting}
\noindent thereafter. However, this step only installs the interfaces but not the actual tools that \textsc{FeynCalc} gets interfaced too.
The only exception is the library of analytic results for Passarino--Veltman functions from \textsc{Package-X}\footnote{The corresponding binary file is directly shipped with \textsc{FeynHelpers} with the kind permission of \textsc{Package-X}'s author Hiren Patel.}. For all other tools we cannot offer a simple installation solution because most of them need to be compiled from the source code on user's machine. 

To simplify this task as much as possible, we recommend using a \textsc{Linux} system, where the build procedures are very much straightforward. \textsc{macOS} is the second best system for this task, but it usually requires the user to install \textsc{homebrew} to fetch the required libraries. In some cases, Apple Silicon might still not be fully supported\footnote{As of now, it appears that KIRA
does not fully support Apple Silicon \cf \url{https://gitlab.com/kira-pyred/kira/-/issues/46}} by all tools interfaced to in  \textsc{FeynHelpers}, which would require {Rosetta} emulation. Finally, \textsc{Windows} is the least suitable operating system for using \textsc{FeynHelpers} in multiloop calculations. For example \textsc{FIRE}, \textsc{FIESTA}, \textsc{KIRA} or \textsc{pysSecDec} completely lack native \textsc{Windows} support and can be at most used via Windows Subsystem for Linux.

In the \textsc{FeynHelpers} manual we tried to summarize some recipes on compiling the interfaced codes based on the instructions provided by the respective developers. However, we would like to emphasize that we do not aim at streamlining the installation of those tools or intend to provide support on compiling the tool X on system Y. Those questions are not related to \textsc{FeynHelpers} or \textsc{FeynCalc} and should be therefore addressed to the developers of the respective software.

Provided that \textsc{FeynCalc} and \textsc{FeynHelpers} as well as all desired programs have been successfully installed, the user still might need to adjust some options the  \textsc{FeynHelpers} routines to ensure that they point to the correct binaries. Let us briefly explain this below.

In \textsc{QGRAF} interface the function \texttt{QGCreateAmp} needs to know the location of the \textsc{QGRAF} binary. The path is specified via the option \texttt{QGBinaryFile}. By default \textsc{FeynHelpers} will look for the binary in \texttt{FileNameJoin[\{\$FeynHelpersDirectory, "ExternalTools", "QGRAF", "Binary"\}]}, where "Binary" is substituted with 
\texttt{"qgraf"} on Linux or macOS. 

As far as \textsc{FIRE} is concerned, the routine \texttt{FIREPrepareStartFile} has an option \texttt{FIREPath} that specifies the location of \textsc{FIRE}'s \textsc{Mathematica} interface. The default value is \texttt{File\-NameJoin[\{\$UserBaseDirectory,\-"Applications", "FIRE7", "FIRE7.m"\}]}. Then, both \texttt{FIRE\-Create\-Start\-File} and \texttt{FIRECreateLiteRedFiles} need to know the full path to a working Mathematica kernel. The default behavior of the corresponding option \texttt{FIRE\-Mathematica\-Kernel\-Path} is to look for it in 
\texttt{FileNameJoin[\{\$InstallationDirectory, "Executables", "math"\}]\} on \textsc{Linux} or in \texttt{FileNameJoin[\{\$InstallationDirectory, "MacOS", "math"}]} on \textsc{macOS}. Finally, \texttt{FIRERunReduction} requires the full path to a FIRE binary, which is specified via the option FIREBinaryPath and set to \texttt{FileNameJoin[\{\$UserBaseDirectory, "Applications", "FIRE7", "bin", "FIRE7"\}]} by default.

In \textsc{KIRA} interface it is only the function \texttt{KiraRunReduction} that needs to know where to find KIRA and FERMAT. The corresponding options are called \texttt{KiraBinaryPath} and \texttt{KiraFermatPath}. The former is set to \texttt{"kira"}, thus assuming that the full path to the binary is already contained in users \texttt{PATH} environmental variable\footnote{Normally the installer of Kira would take care of it.}. The latter points to  FileNameJoin[\{\$FeynHelpersDirectory, "ExternalTools", "Fermat", "ferl6", "fer64"\}] on \textsc{Linux} and FileNameJoin[\{\$FeynHelpersDirectory, "ExternalTools", "Fermat", "ferm6", "fer64"\}] on \textsc{macOS}.

To use \textsc{FIESTA}, the function \texttt{FSACreateMathematicaScripts} needs to know the location of the Mathematica package. The default value of the corresponding option \texttt{FSAPath} is \texttt{FileNameJoin[\{\$UserBaseDirectory, "Applications", "FIESTA5", "FIESTA5.m"\}]}. In addition to that,  \texttt{FSARunIntegration} has an option \texttt{FSAMathematicaKernelPath} that points to a Mathematica kernel. The default values are the same as in the case of FIRE's \texttt{FIRE\-Mathematica\-KernelPath} option.

For \textsc{LoopTools} we need to specify the full path to the \textsc{MathLink} binary of the library via the option \texttt{LToolsPath} of \texttt{LToolsLoadLibrary}. The default value here is  \texttt{FileNameJoin[\ \{\$FeynHelpers\-Directory, "ExternalTools", "LoopTools", "LoopTools"\}]}.

Finally, the usage of \textsc{FERMAT} requires setting the option \texttt{FerPath}, which has the same default values as \textsc{KIRA}'s
\texttt{KiraFermatPath}.

\section{Documentation} \label{sec:docu}

We intend to keep the description of different interfaces succinct, as \textsc{FeynHelpers} comes with an extensive, almost 200 pages long documentation covering every symbol, function and option available in the add-on. The documentation is available both in form of a PDF\footnote{\url{https://github.com/FeynCalc/feynhelpers-manual/releases/tag/dev-manual}}  and online\footnote{\url{https://feyncalc.github.io/referenceFeynHelpersDev}}. It also contains a short tutorial for each interface that should help the user to get things running swiftly. 

\section{Interfaces} \label{sec:interfaces}

\subsection{QGRAF} \label{sec:int-qgraf}

\texttt{QGRAF} is a high-performance Feynman diagram generator widely employed in the multiloop community. The code uses \emph{input files} that specify the process to generate and possibly some additional constraints, \emph{model files} that describe the allowed interactions and \emph{style files} that customize the output format.  Unlike \textsc{FeynArts} that requires models with built-in
Feynman rules and can plot Feynman diagrams directly in \textsc{Mathematica} notebooks, \textsc{QGRAF} can only output diagrams in form of text files using conventions provided by the style files. Typically, such output consists of placeholders representing propagators, vertices and external states of  the given model. Replacing these placeholders with explicit Feynman rules is not part of \textsc{QGRAF} and is essentially left to the user.
 
For the reasons given above, generation of Feynman diagrams using \textsc{QGRAF} is always a multi-step process. To invoke \textsc{QGRAF} via \textsc{FeynHelpers} the user needs to provide a valid model file and specify some process involving particles present in that model. The interface already contains suitable style files for producing \textsc{Mathematica}-readable expressions and drawing the diagrams. Custom style files are supported as well. 

In the default output style file \texttt{feyncalc.sty} external states are denoted as \texttt{QG\-Polar\-ization[$\psi$[$i$,$p$],$m_\psi$]}, where $\psi$ stands for a field with momentum $p$ and mass $m_\psi$, while $i$ is the number of the node connecting this edge to the rest of the diagram. Accordingly, we use  \texttt{QGPropagator[$\psi$[$i$,$p$],$\bar{\psi}$[$j$,$-p$], $m_\psi$]} for propagators and \texttt{QGVertex[$\psi_1$[$i_1$,$p_1$], $\psi_1$[$i_2$,$p_2$], $\ldots$ ]} for vertices. Notice that node numbers are used to reconstruct all contractions of various indices (\eg Lorentz, Dirac, color etc.) appearing in amplitude.

To generate an amplitude one should use the function \texttt{QGCreateAmp} that accepts two arguments. The first one is the number of loops, while the second one is a list describing the process we want to generate. The syntax of the  latter is

\begin{lstlisting}[extendedchars=true,language=Mathematica]
    {{InParticle1[p1], InParticle2[p2], ...}}->{OutParticle1[q1],OutParticle2[q2],...}
\end{lstlisting}
\noindent where the particle names are defined by the \textsc{QGRAF} model in use\footnote{
For the built-in models we use a naming convention described in \texttt{FileNameJoin[\{\$FeynCalcDirectory, "Examples"\}]\}}}. The model can be specified via the \texttt{QGModel} option. Its value can be either one of the built-in models\footnote{The built-in models are located in \texttt{FileNameJoin[\{\$FeynHelpersDirectory, "ExternalTools", "QGRAF", "Models"\}]\}}. Currently those include pure QED and QCD only.} or the full path to a user-defined model.
The naming scheme for the loop momenta can be fixed using the option \texttt{QGLoopMomentum}. Setting it \eg to \texttt{l} means that the loop momenta will be named \texttt{l1}, \texttt{l2}, \texttt{l3} etc. If one wants to pass some options directly to \textsc{QGRAF},
this can be done via \texttt{QGOptions}. For example, to enforce the commonly used restricitons for filtering out loop corrections on external legs and tadpole subdiagrams, one can set this option to \texttt{\{"notadpole","onshell"\}}. Last but not least, one should also specify the output directory to store the generated amplitudes. In the following, we assume it to be \texttt{dir}. Then, an example code to generate one-loop diagrams for the process $e^- (p_1) e^+ (p_2) \to  e^- (p_3) e^+ (p_4)$ using the built-in one-flavor QED model would be
\begin{lstlisting}[extendedchars=true,language=Mathematica]
    qgOutput=QGCreateAmp[1,{"El[p1]","Ael[p2]"}->{"El[p3]","Ael[p4]"},QGModel->"QEDOneFlavor",
    QGLoopMomentum->l,QGOptions->{"notadpole","onshell"}, QGOutputDirectory->dir];
\end{lstlisting}
\noindent The next step is to load a set of Feynman rules needed for the model. For the built-in models it is sufficient to evaluate \texttt{QGLoadInsertions["QGCommonInsertions.m"]}. Of course, one can also use a custom set of rules designed for the specific model. In this case the function argument must be the full path to that file. The Feynman rules are written as replacement rules in a list, where the right hand side of each rule contains a \texttt{QGPolarization}, \texttt{QGVertex} or \texttt{QGPropagator} symbol with placeholder arguments matching expressions that arise in the given model\footnote{The file \texttt{QGCommonInsertions.m} located in  \texttt{FileNameJoin[\{\$FeynHelpersDirectory,"ExternalTools", "QGRAF","Insertions"\}]\}} can be used as a blueprint for implementing custom Feynman rules.}.

The insertion of Feynman rules into \textsc{QGRAF} amplitudes can be accomplished via
\begin{lstlisting}[extendedchars=true,language=Mathematica]
    qgAmps = QGConvertToFC[qgOutput];
\end{lstlisting}
\noindent where \texttt{qgAmps} is a list of amplitudes ready to be evaluated in \textsc{FeynCalc}. Notice that the insertion files to be used by this function are determined by the option \texttt{QGInsertionRule}. Its default value is \texttt{\{"QGCommonInsertions"\}}.

There are two popular ways to visualize \textsc{QGRAF} output. The first one is to use \textsc{Graphviz}, which allows for a fast, albeit somewhat low-quality, graphical representation of thousands of diagrams. Publication-quality images  can be created using \textsc{TikZ-Feynman} via \textsc{LuaTeX}. However, in this case the compilation times are much longer as compared to \textsc{GraphViz}, because \textsc{LuaTeX} is not optimized for fast output generation. Therefore, even when parallelized it is often still too slow to visualize thousands or more diagrams within a reasonable time frame.

Nevertheless, \textsc{FeynHelpers} explicitly supports diagram generation with \textsc{TikZ-Feyn\-man} by creating the corresponding tex file when running 
\texttt{QGCreateAmp} and providing a routine that helps to define the styling\footnote{Without the styling every particle will be represented as a solid line.} of the involved fields. In particular, in the above example of generating QED diagrams, one could evaluate 
\begin{lstlisting}[extendedchars=true,language=Mathematica]
    tikzStyles=QGTZFCreateFieldStyles["QEDOneFlavor", qgOutput,
    QGFieldStyles->{{"Ga","photon","\\gamma"}, {"El","fermion","e\^{}-"}, 
        {"Ael","anti fermion","e\^{}+"}}]}
\end{lstlisting}
\noindent to create a styling file \texttt{tikz-styles.tex}.
The field line type and label (second and third elements of each list) need to be provided by hand, as this information is needed to improve the \textsc{TikZ-Feynman} output and is not contained in the QGRAF model. After that one can run
\begin{lstlisting}[extendedchars=true,language=Mathematica]
    QGTZFCreateTeXFiles[qgOutput,Split->True]
\end{lstlisting}
\noindent to generate a separate tex file for each diagram. Each of these files can be then directly compiled with \textsc{LuaTeX}. On \textsc{Linux} or \textsc{macOS} one could also use the automatically generated scripts \texttt{makeDiagrams.sh} and \texttt{glueDiagrams.sh} that employ \textsc{GNU parallel}~\cite{tange_2020_3996295} and \textsc{pdfunite} utilities to compile the files in parallel and then merge them into a single PDF file. Notice that \texttt{QGTZFCreateFieldStyles} needs to be evaluated only once for each new model, since \texttt{QGTZFCreateTeXFiles} can then use the same \texttt{tikz-styles.tex} for any process within this model.

\subsection{FIRE} \label{sec:int-fire}

\textsc{FIRE} is a program for automatic IBP-reduction of multiloop integrals that is written in \texttt{Mathematica} and \texttt{C++}. The \texttt{Mathematica} part is used to analyze the supplied integral family and generate the so-called \emph{start files} needed for the reduction. The quality of this analysis can be improved (\eg to detect more symmetries between different sectors) by using \textsc{LiteRed}, which is also recommended by the developers. Apart from the start files one also needs a file containing integrals to be reduced and a \emph{config file} specifying how to run the reduction and where to save the results. The reduction itself is done topology-wise by running the \texttt{C++} engine of \textsc{FIRE}. Its output consists of a \emph{tables file} containing reduction tables for the given list of integrals. The so-obtained set of master integrals is, however, not guaranteed to be minimal, as there still might be one-to-one relations between some of the masters\footnote{These mappings can be revealed \eg using \textsc{FeynCalc}'s function \texttt{FCLoopFindIntegralMappings}.}.

The \textsc{FIRE} interface of \textsc{FeynHelpers} allows for automatizing all steps necessary to set up the reduction, where the user only needs to provide a list of integrals (in the \texttt{GLI}-notation) and topologies (in the \texttt{FCTopology}-notation). In the following we will denote these two lists as \texttt{ints} and \texttt{topos} respectively.

The first step is to evaluate
\begin{lstlisting}[extendedchars=true,language=Mathematica]
   FIREPrepareStartFile[topos,dir];
\end{lstlisting}
\noindent which will create a directory named after the topology identifier inside \texttt{dir} and populate it with two \textsc{Mathematica} scripts: \texttt{CreateLiteRedFiles.m} and \texttt{CreateStartFile.m}. The first script uses \textsc{LiteRed} to analyze the given topology and save the results in a format that
can be understood by \textsc{FIRE}. Then, the second script creates \textsc{FIRE} start files using the results from \textsc{LiteRed}. For complicated topologies with many scales and propagators the evaluation times for these two scripts can easily go into many minutes or even hours. This is why it is recommended to run them on the same workstation or computer cluster where the main reduction is being carried out. For simple cases (\eg two-loop families with one scale) it is also possible to complete this step without leaving the notebook by starting a separate \textsc{Mathematica} instance in the background. This is done by issuing
\begin{lstlisting}[extendedchars=true,language=Mathematica]
    FIRECreateLiteRedFiles[dir,topos];
    FIRECreateStartFile[dir,topos];
\end{lstlisting}
The next step is to generate the config files with
\begin{lstlisting}[extendedchars=true,language=Mathematica]
    FIRECreateConfigFile[topos,dir];
\end{lstlisting}
\noindent By default, this routine assumes that the reduction will be done on the current machine and adjusts the number of the simultaneous thread according to the number of present CPU cores. For the more realistic case of doing the reduction on a cluster, one can tweak the corresponding settings using the options \texttt{FIREFthreads}, \texttt{FIRESthreads}, \texttt{FIRELthreads} and \texttt{FIREThreads}.
The list of integrals to be reduced is added to \texttt{dir} with the command
\begin{lstlisting}[extendedchars=true,language=Mathematica]
    FIRECreateIntegralFile[ints,topos,dir];
\end{lstlisting} 
\noindent After that, the reduction is ready to be started. Again, for very simple cases with only few integral families, where each of them requires only some seconds to finish, one can initiate the reduction directly from the notebook via 
\begin{lstlisting}[extendedchars=true,language=Mathematica]
    FIRERunReduction[dir,topos];
\end{lstlisting} 

Provided that the reduction finished successfully, one still needs to load the reduction tables and convert them to a set of reduction rules for \texttt{GLI}s in the unreduced amplitude. This can be conveniently done using
\begin{lstlisting}[extendedchars=true,language=Mathematica]
    tables=FIREImportResults[topos,dir];
\end{lstlisting} 
\noindent with \texttt{tables} containing lists of reduction rules for all topologies from  \texttt{topos}. At this point we can insert those rules into the original amplitude, thus completing this step of the computation.

Notice that the interface does not yet support the finite field mode of \textsc{FIRE}, which might be required for some particularly complicated reductions with multiple kinematic invariants.

\subsection{KIRA} \label{sec:int-kira}

\textsc{KIRA} is another popular tool for doing IBP reductions. The program is written in \textsc{C++} and is steered using \textsc{YAML} files. There are \emph{config files} that describe  integral families and kinematic constrains as well as \emph{job files}, which tell \textsc{KIRA} what to do during the run.

Automatized IBP reduction with \textsc{KIRA} is equally easy to set up as in the case of \textsc{FIRE}. Again we start with our list of \texttt{GLI}-integrals (\texttt{ints}) and \texttt{FCTopology}-topologies (\texttt{topos}), while the reduction input and output will be located in a subdirectory of \texttt{dir}.

First of all, we need to generate a job file, which can be done with 

\begin{lstlisting}[extendedchars=true,language=Mathematica];
    KiraCreateJobFile[topos,ints,dir];
\end{lstlisting} 
\noindent Normally, \textsc{KIRA} requires the user to specify explicitly, which sectors of the given integral family need to be reduced. While doing this manually can be useful to fine-tune the reduction process, in \textsc{FeynHelpers} we adapt a more pragmatic and user-friendly approach by letting the code analyze the provided list of loop integrals and automatically select their top sectors\footnote{Notice that it is also possible to specify the sectors manually using $r,s$-notation of \textsc{KIRA}.}. 

Having successfully created job files for all topologies, one still needs to export the list of integrals to be reduced. This merely requires 
\begin{lstlisting}[extendedchars=true,language=Mathematica];
    KiraCreateIntegralFile[ints,topos,dir];
\end{lstlisting} 
The last step of setting up the reduction is the creation of the config files using
\begin{lstlisting}[extendedchars=true,language=Mathematica]
    KiraCreateConfigFiles[topos,ints,dir, KiraMassDimension -> {...}];
\end{lstlisting} 
\noindent Notice that the option \texttt{KiraMassDimension} is compulsory, as \texttt{KIRA} needs to know the mass dimensions of all kinematic invariants appearing in the integral topology. For example, if the reduction depends on a squared momentum \texttt{pp} and a mass \texttt{m}, then we must set this option to \texttt{\{pp->2, m-> 1\}}.

After these preparatory steps we are ready to run the reduction with \textsc{KIRA}. Just as in the case of \textsc{FIRE}, we provide a convenience function for starting the reduction from the current notebook. At the same time, we highly discourage the users from using it, unless it is obvious that the process can finish within a couple of minutes. In the \textsc{KIRA} interface this routine is called
\begin{lstlisting}[extendedchars=true,language=Mathematica]
    KIRARunReduction[dir,topo];
\end{lstlisting} 

Finally, once all reductions finished one can conveniently load the reduction rules via
\begin{lstlisting}[extendedchars=true,language=Mathematica]
    tables=KiraImportResults[topos,dir];
\end{lstlisting} 

The \textsc{KIRA} interface does not yet support reductions using finite field, but it is planned to be added in the future.

\subsection{pySecDec} \label{sec:int-psd}

\textsc{pySecDec} provides a convenient way to evaluate multiloop integrals numerically at different kinematic points using sector decomposition~\cite{Binoth:2000ps}. While one can use a \textsc{Python} front-end to set up the calculation, the computationally expensive routines are implemented in \textsc{FORM} and \textsc{C++}. The common way to handle an integral with 
\textsc{pySecDec} consists of preparing two \textsc{Python} scripts, which we call \texttt{generate.py} and \texttt{integrate.py}. The former contains the definition of the integral, \eg its  propagators and their powers, replacement rules for kinematic invariants etc. Evaluating this script invokes \textsc{pySecDec}'s routines to analyze the integral and separate its singularities from the finite part by means of sector decomposition. The output of this analysis is saved to a user-specified directory (\eg \texttt{loopint}) that also contains a \texttt{Makefile}. By running something like
\begin{lstlisting}[extendedchars=false,language=Bash]
    make -C loopint -j8
\end{lstlisting} 
\noindent one can compile the code to a library in parallel using 8 threads. Notice that before compiling the \textsc{C++} code, the script will call \textsc{FORM} to do  symbolical manipulations of the integrand required by the sector decomposition approach. After that one can run the script \texttt{integrate.py}, which calls the \textsc{C++} library from \texttt{loopint} for the desired values of kinematic invariants. Furthermore, at this stage the user can still choose between several built-in integrators and change their options.

The \textsc{FeynHelpers} interface to \textsc{pySecDec} focuses on creating \texttt{generate\-.py} and \texttt{in\-tegrate.py} for the given GLI-integrals \texttt{ints} belonging to the \texttt{FCTopology}-families \texttt{topos}. All files related to a particular integral will be placed to a separate directory inside \texttt{dir}. The directory name is generated from the integral family identifier and the present propagator powers.

Furthermore, it is important to specify some important options, such as values of the kinematic invariants and the desired expansion in the dimensional regulator $\varepsilon$. Depending on whether these values are real or complex numbers one should use \texttt{PSD\-Real\-Parameter\-Rules} or \texttt{PSD\-Complex\-Parameter\-Rules} respectively. Both options are lists of the form \texttt{\{invariant -> value, \ldots \}}. By default, each integral is only calculated up to
$\mathcal(\varepsilon^0)$. To change this behavior one should use the option \texttt{PSDRequestedOrder}.

A typical command to generate both \textsc{pySecDec} scripts would look like

\begin{lstlisting}[extendedchars=true,language=Mathematica]
    files=PSDCreatePythonScripts[ints, topos, dir, PSDRealParameterRules -> {...}, PSDComplexParameterRules -> {...}, PSDRequestedOrder->2];
\end{lstlisting} 
\noindent Contour deformation, that is crucial to calculate integrals with imaginary parts, is turned on by default. However, if one knows that the integral in question is purely real, turning off contour deformation will make the evaluation much faster. This can be done via the option \texttt{PSDContourDeformation}. Let us also remark that trying to run the same command again would generate an error message, because this would overwrite the existing script files. To allow for this one must set the option
\texttt{OverwriteTarget} to \texttt{True}. Furthermore, \texttt{PSDCreatePythonScripts} outputs a list (which we call \texttt{files}) containing full paths to the \texttt{generate.py} and \texttt{integrate.py} scripts for each loop integral. This list will be needed at a later stage as we will explain below.

For the  time being we do not provide a dedicated function for evaluating an integral with \textsc{pySecDec} directly from a notebook. The reason for this is that the runtimes of \texttt{generate.py} and \texttt{make} can be substantial, even for integrals that appear simple and involve only a small number of edges. This is because such integrals may have bad divergences that require multiple sector decompositions to be regularized properly. Furthermore, the integration process may easily run for hours or even days if the integration at the given kinematic point is converging too slowly. Therefore, we believe that it if the evaluation is to be done on the user's machine,
it is better to directly control the process and take measures in case of issues.

Upon finishing the integration, the \texttt{integrate.py} script will save the numerical results for the given integral to a \textsc{Mathematica}-readable file of the form \texttt{numres\-\_val1\-\_val2\-\_...\_mma.m}, where each file contains a list of two elements, the first being the numerical value and the second describing the uncertainty. To facilitate the task of loading these numerical values into the current notebook, \textsc{FeynHelpers} offers  a mechanism, where the user just needs to provide the \texttt{files}-list and the values of kinematic invariants that were chosen when generating the scripts. In this case running
\begin{lstlisting}[extendedchars=true,language=Mathematica]
    res=PSDLoadNumericalResults[files, PSDRealParameterRules -> {...}, PSDComplexParameterRules -> {...}];
\end{lstlisting}
\noindent will automatically load the correct files. After that it is enough to evaluate something like
\begin{lstlisting}[extendedchars=true,language=Mathematica]
    repRule = Thread[Rule[ints,First/@res]];
\end{lstlisting}
to obtain a list of replacement rules assigning numerical values to all loop integrals from the \texttt{ints} list. 

It is worth noting that the interface also supports asymptotic expansions~\cite{Beneke:1997zp} that were added to \textsc{pySecDec} several years ago~\cite{Heinrich:2021dbf}. To activate this mode it is sufficient to specify the small parameter via 
\texttt{PSD\-Expansion\-By\-Regions\-Parameter} and use the option \texttt{PSDExpansion\-By\-Regions\-Order} to set the expansion order. 
However, it is still necessary to assign a numerical value to the small parameter via \texttt{PSD\-Real\-Parameter\-Rules} or
\texttt{PSD\-Complex\-Parameter\-Rules}. Unlike \textsc{FIESTA}, \textsc{pySecDec} currently cannot return the result of an asymptotic expansion in a form, where the small parameter remains symbolic.

\subsection{FIESTA} \label{sec:int-fiesta}

\textsc{FIESTA} offers an alternative to \textsc{pySecDec} when it comes to numerical evaluation of multiloop integrals. Both codes rely on sector decomposition, but implement it in a different way so that one can use them for cross checks of numerical results. Since \textsc{FIESTA}'s front end is written in \textsc{Mathematica}, interfacing it with other \textsc{Mathematica} codes is somewhat easier as compared to \textsc{pySecDec}

When it comes to creating run cards for \textsc{FIESTA}, there is an important caveat related to the sign of $i \eta$ in the propagators. As was explained in Sec.~4.2 of ref.~\cite{Shtabovenko:2023idz}, in \textsc{FIESTA} the entered propagator denominators
are understood to contain an implicit $-i \eta$, contrary to the convention used in \textsc{FeynCalc} or \textsc{pySecDec}, where it is a $+i \eta$ . Hence, if we want to evaluate an integral containing $[p^2 -m^2 + i \eta]^{-1}$, we have to first rewrite this denominator as
$-[-p^2 +m^2 - i \eta]^{-1}$ and then pass ${m^2-p^2}$ to \textsc{FIESTA}, keeping track of the overall sign in front of the integral.
To automatize this procedure as much as possible, \textsc{FeynCalc} has a built-in routine \texttt{FCLoopSwitchEtaSign} that can be used to convert \texttt{FCTopology} containers to \textsc{FIESTA}'s sign convention. Given a list of \texttt{GLI}-integrals (\texttt{ints}) and \texttt{FCTopology}-topologies (\texttt{toposRaw}) it is sufficient to evaluate 
\begin{lstlisting}[extendedchars=true,language=Mathematica]
    topos = FCLoopSwitchEtaSign[toposRaw,-1];
\end{lstlisting}
The resulting list of topologies \texttt{topos} can be then directly passed to \textsc{FIESTA}.

The interface to \texttt{FIESTA} essentially generates a \textsc{Mathematica} script file \texttt{FiestaScript.m} for each integral. This script goes to a separate directory inside \texttt{dir}, where the directory name is made of family identifier and propagator powers.
The script then loads \textsc{FIESTA} package, calculates the integral and saves the result to a file of the form  \texttt{numres\-\_val1\-\_val2\-\_...\_fiesta.m}

To generate such script one needs to run the function \texttt{FSACreate\-Mathematica\-Scripts}, specifying numerical values of kinematic invariants (as usual replacement rules) via \texttt{FSA\-Parameter\-Rules} and the desired order in $\varepsilon$ using \texttt{FSAOrderInEps}. An example would be
\begin{lstlisting}[extendedchars=true,language=Mathematica]
    files = FSACreateMathematicaScripts[ints, topos, dir, FSAParameterRules -> {...}];
\end{lstlisting}
where, as in the case of the \textsc{pySecDec}-interface, any attempt to overwrite the existing script file will lead to an error,
unless the option \texttt{OverwriteTarget} is set to \texttt{True}. If needed, the contour deformation can be deactivated by setting 
\texttt{FSAComplexMode} to \texttt{False}.

The output \texttt{files} is a list of two-element lists containing the full path to the \textsc{Mathematica} script file \texttt{FiestaScript.m} and to the output file with the numerical result. For sufficiently simple integrals one can evaluate the scripts directly by running 
\begin{lstlisting}[extendedchars=true,language=Mathematica]
    FSARunIntegration[First/@files];
\end{lstlisting}
in the current notebook. At this point we would like to warn the user, that \textsc{FIESTA} may sometimes issue warning and error messages during the integration but nevertheless output some (possibly) incorrect result. This is why running evaluation scripts blindly via \texttt{FSA\-Run\-Integration}, where one cannot see such messages, can be risky.

In any case, once the numerical results are there, they can be loaded via 
\begin{lstlisting}[extendedchars=true,language=Mathematica]
    FSALoadNumericalResults[files];
    repRule = Thread[Rule[ints,First/@res]];
\end{lstlisting}
where \texttt{repRule} contains replacement rules for \texttt{GLI} integrals.

If needed, the interface can be also used to perform asymptotic expansions in small parameters. To that aim
one needs to set the option \texttt{FSASDExpandAsy} to \texttt{True}, and specify the expansion
parameter and the expansion order using \texttt{FSAExpandVar} and \texttt{FSASDExpandAsyOrder} respectively.

\subsection{LoopTools} \label{sec:int-lt}

\textsc{LoopTools} is one of numerous codes (\cf \eg refs.~\cite{Hahn:1998yk,Denner:2016kdg,Denner:2002ii,Denner:2005nn,Denner:2010tr,Ellis:2007qk,vanHameren:2010cp}) for numerical evaluation of Passarino--Veltman~\cite{Passarino:1978jh} functions. Given the facts that it already has a \textsc{Mathematica} interface and uses the same momentum-routing conventions as \textsc{FeynCalc}, adding a \textsc{LoopTools} interface to \textsc{FeynHelpers} was comparably simple task.

To get it running one merely needs to load the \textsc{MathLink} interface to \textsc{LoopTools} on the current kernel using
\texttt{LToolsLoadLibrary[]}. Notice that this command should be evaluated only once per \textsc{Mathematica} session. After that, to compute all supported \texttt{PaVe}-functions in \texttt{exp} numerically, it is sufficient to run
\begin{lstlisting}[extendedchars=true,language=Mathematica]
    LToolsEvaluate[exp, LToolsSetMudim -> 1, InitialSubstitutions -> {invariant->val, ...}]
\end{lstlisting}
Here \texttt{LToolsSetMudim} specifies the value of the renormalization scale $\mu$, while \texttt{Initial\-Substitutions} contains numerical replacement rules
for the kinematic invariants.

Let us also clarify some subtleties regarding the normalization. In \texttt{FeynCalc}, \texttt{PaVe}-functions are defined to have an overall $1/(i \pi)^2$ prefactor, while \textsc{LoopTools} uses $(i \pi^{D/2} r_\Gamma)^{-1}$ with $D = 4- 2 \varepsilon$ and $r_\Gamma = \Gamma^2(1-\varepsilon) \Gamma(1+\varepsilon)/ \Gamma(1-2\varepsilon)$. By default, \texttt{LToolsEvaluate} will return the full result up to $\mathcal{O}(\varepsilon^0)$ and account for the difference between the prefactors by multiplying the output with $r_\Gamma/\pi^\varepsilon$. Hence, the final result will be in agreement with the \textsc{FeynCalc} convention. However, if the function is requested to return a particular coefficient of $\varepsilon$ by setting the option  \texttt{LToolsFullResult} to \texttt{False}, then there will be no compensating prefactor and the user will receive the direct output of \textsc{LoopTools} using \textsc{LoopTools} normalization.

\subsection{FERMAT} \label{sec:int-fermat}

\texttt{FERMAT}~\cite{Lewis:Fermat} is a computer algebra system that features very efficient algorithms for symbolic manipulations of polynomials and matrices. In particular, the program is used in many multiloop codes such as \texttt{FIRE}, \texttt{LiteRed}, \texttt{Libra}~\cite{Lee:2020zfb} or \texttt{KIRA}, where it is important to have fast routines for factoring multivariate polynomials.

In the current version of \textsc{FeynHelpers} we provide an interface only to a very small subset of \texttt{FERMAT}'s capabilities, namely its solver for symbolic systems of linear equations. The corresponding function is called \texttt{FerSolve} and has syntax similar to \textsc{Mathematica}'s \texttt{Solve}, in the sense that its first argument is a list of linear equation, while the second argument specifies variables this system should be solved for.

In the background, \texttt{FerSolve} generates a  \texttt{FERMAT} script for the given system, runs it and parses the result back to 
\textsc{Mathematica}. A simple example of using the function would be

\begin{lstlisting}[extendedchars=true,language=Mathematica]
    FerSolve[{2 x + y c == 2, 4 x == c}, {x, y}]
\end{lstlisting}

While \texttt{FerSolve} is slower than \texttt{Solve} for simple systems of equations due to the scripting overhead, it easily outperforms \textsc{Mathematica} on larger systems (\eg $10 \times 10$ and beyond, especially when those are dense). One of the immediate applications for this functionality related to \textsc{FeynCalc} are tensor reductions for integrals with high tensor rank or many legs. The \textsc{FeynCalc} routine for deriving such decompositions can use \texttt{FerSolve} instead of \texttt{Solve} when determining the coefficients multiplying tensor basis elements. For example, in the following decomposition of a rank-6 two-loop tensor integral
with two external legs

\begin{lstlisting}[extendedchars=true,language=Mathematica]
    Tdec[{{k1, mu1}, {k1, mu2}, {k1, mu3}, {k1, mu4}, {k2, mu5}, {k2, mu6}},
        {q1, q2}, Solve -> FerSolve, UseTIDL -> False, FCVerbose -> 1]
\end{lstlisting}

we end up with a $52 \times 52$ symbolic matrix describing the resulting system of equations. On a modern laptop \texttt{FERMAT} can solve it within 40 seconds, while both \texttt{Solve} and \textsc{FeynCalc}'s \texttt{Solve3} need multiple minutes to derive a solution.

\section{News from FeynCalc} \label{sec:fc}

\subsection{Parallelization} \label{sec:fc-parallel}

Although \texttt{Mathematica} supports parallel calculations since 2008, parallelizing existing codes often turns out to be challenging. This is related to the fact that essential routines such as \texttt{Simplify}, \texttt{Factor}, \texttt{Expand}, \texttt{Together}, \texttt{Series} or \texttt{Solve} cannot seamlessly make use of multicore CPUs. Instead, the user has to carefully split the expressions to be evaluated into smaller chunks and distribute them over parallel \textsc{Mathematica} kernels. Every kernel needs to be provided with data and code required to evaluate the given expressions. Finally, once the calculations are done, parallel kernels will return the results to the main kernel
where the user can access them. Trying to accomplish this task by naively applying \texttt{ParallelMap} or \texttt{ParallelTable} to some single-threaded
codes often not only fails to achieve any measurable performance boost but even makes the evaluation take longer as compared to doing everything on the same kernel. The situation becomes even more complicated when trying to parallelize large and complicated packages such as \textsc{FeynCalc}, with almost one thousand symbols and functions as well as multiple values dynamically defined during the runtime. Here it is important not only to ensure that the code can run efficiently when distributed to parallel kernels but also to avoid cases, where \eg definitions of scalar products or noncommutative variables differ between different kernels. 

For these reasons the issue of parallel evaluation was not addressed in \textsc{FeynCalc} for long time. However, with the introduction of multiloop-related functionality in version 10, it became obvious that some operations (\eg deriving characteristic polynomials for thousands of integral families or finding mappings between tens of thousands of master integrals) are too slow for realistic multiloop calculations, while they could highly benefit from parallel evaluation. After some trial and error we managed to come up with a good pattern for parallelizing different \textsc{FeynCalc} functions, that allows us to successively introduce parallel evaluation for selected pieces of the package. Let us now explain how this works in details.

Having loaded \textsc{FeynCalc} on a fresh kernel, the first step is to start some parallel subkernels controlled by the main kernel. Then we need to enable the parallelization of the package via the global variable \texttt{\$ParallelizeFeynCalc}. This will load \textsc{FeynCalc} on each of the parallel kernels, ensuring that all \textsc{FeynCalc}-related symbols and functions are available for parallel evaluation. An example code for this would be

\begin{lstlisting}[extendedchars=true,language=Mathematica]
    <<FeynCalc`
    LaunchKernels[8];
    $ParallelizeFeynCalc = True;
\end{lstlisting}
Let us stress, that even after these steps the parallelization of each \textsc{FeynCalc} function must be enabled manually.
To this aim each routine with support for parallel evaluation has an option \texttt{FCParallelize} set to \texttt{False} by default.
The only exception to this rule are \texttt{ScalarProduct}, \texttt{DataType}, \texttt{FCClearScalarProducts}, \texttt{FCClearDataTypes} and the functions for defining noncommutative quantities\footnote{
These include \texttt{(Un)DeclareNonCommutative}, \texttt{FC(Anti)Commutator} and \texttt{(Un)Declare(Anti)Commutator.}
}. The reason for this is that every \texttt{ScalarProduct}, \texttt{DataType} or noncommutative definition necessarily must be propagated to all parallel kernels to avoid inconsistencies in the calculation. This is why we recommend avoiding cases where the user first does some calculations with \textsc{FeynCalc} in the single-threaded mode and then decides to activate the parallel evaluation at a later point.
Ideally, the parallelization should be enabled from the start on a fresh kernel.

Apart from the above-mentioned special functions, all other routines supporting parallelization will activate the multi-threaded mode 
only if \texttt{FCParallelize} is set to \texttt{True}. Naively, one might think that it would have been more practical to enable parallelization everywhere by default. However, if done without any precautions this might cause problems, as parallelizable functions may call other parallelizable functions and since every parallelization involves distributing tasks to parallel kernels, we must avoid the situation of trying to parallelize something that already runs on a subkernel. Since parallel evaluation is a rather new concept in using \textsc{FeynCalc} for calculations, we might adjust the default behavior in the future after receiving more user feedback on this matter.

In the current implementation most parallelizable functions expect a list of expressions (\eg Feynman diagrams for \texttt{Contract}, \texttt{SUNSimplify} or \texttt{DiracSimplify}) that they evaluate on parallel kernels. For multiloop-related functions such as \texttt{FCLoopFindTopologyMappings}, \texttt{FCLoopFindIntegralMappings} or \texttt{FCFeynmanPrepare} such input could be a list of integrals or topologies. Multi-threaded evaluation of a single expression (\eg trace or integral) is more challenging, because not every algorithm can be easily modified to be parallelizable\footnote{Algorithms that process large lists using \texttt{Map} or \texttt{Table}
    are good candidates for effortless parallel evaluation. Calculational steps that involve the whole expression (\eg when using \texttt{Expand} or \texttt{Factor}) are difficult to parallelize without redesigning the whole code logic.}. As of now, only \texttt{Collect2} supports such parallel evaluation mode.

While the parallelization is currently enabled only for a limited subset of \textsc{FeynCalc} functions, we plan to add this feature to more functions in the upcoming versions of the package. Nevertheless, many of the examples shipped with \textsc{FeynCalc} have already been updated to make use of parallel kernels, which involves not only loop but also tree-level calculations. Last but not least, in Table \ref{tab:benchmarks} we present several benchmarks corroborating that parallel evaluation does help to speed up \textsc{FeynCalc} calculations on multi-core CPUs. Of course, in practice the results will vary depending on the current calculation and its bottlenecks.

\begin{table}
    \centering
\begin{tabular}{|l|c|c|}
    \hline
    Function &  Single kernel, time in s  & 8 parallel kernels, time in s  \\
    \hline
    \texttt{Contract} & 15.9 & 3.1  \\
    \hline
    \texttt{DiracSimplify}  & 27.0 &  7.1 \\
    \hline
    \texttt{SUNSimplify}  & 24.7 & 8.8 \\
    \hline
\end{tabular}
\caption{FeynCalc performance on a laptop equipped with 8-core AMD Ryzen 7 PRO 4750U and 32 GB RAM running Mathematica 14.3 on Fedora
    Linux when evaluating 21 two-loop diagrams for $g g \to H$}
    \label{tab:benchmarks}
\end{table}

\subsection{Tensor reduction for zero Gram determinants} \label{sec:fc-zerogram}

Tensor integrals are ubiquitous in loop calculations and their treatment often happens to be the last step before setting up IBP reduction and arriving at a finite set of scalar master integrals. Two common approaches to this problem are known under the name of projector techniques and tensor reduction. In the former case one constructs a set of projectors suited for the given process that map\footnote{The mapping often involve calculations of Dirac and color traces.} tensor integrals in the amplitude to scalar ones. Tensor reduction goes back to the idea~\cite{Melrose:1965kb,Passarino:1978jh} that a tensor integral can be decomposed into a linear combination of scalar integrals multiplying suitable tensor structures made of metric tensors and external momenta. Over the years, both projectors and tensor reduction technologies have been developed further to reduce the amount of computational effort and optimize the complexity of the output \cf \eg~\cite{Davydychev:1991va,Binoth:2008uq,Diakonidis:2009fx,Fleischer:2011nt,Denner:2005nn,Chen:2019wyb,Peraro:2019cjj,Hu:2021nia,Chen:2024xwt,Goode:2024cfy,Reeck:2024iwk}. In particular, we refer to ref.~\cite{Goode:2024cfy} for a detailed list of related publications.

Nevertheless, the traditional reduction in the spirit of Passarino and Veltman, where the tensor basis is constructed using momenta 
occurring in the tensor integral denominators and the scalar coefficients are obtained from solving the resulting linear system of equations still remains very popular in practical calculations at one loop and beyond. The reason for this is that the underlying algorithm is very easy to understand and implement, while its technical drawbacks\footnote{The standard approach suffers from numerical instabilities when inverse Gram determinants approach zero and leads to an extreme proliferation for more complicated integrals.} do not necessarily affect every calculation. The amount of tensor basis elements can be significantly reduced by cleverly exploiting the symmetries of the integral as explained in ref.~\cite{Pak:2011xt}, while using a powerful solver such as \textsc{FERMAT}( \cf Sec.~\ref{sec:int-fermat}) or \textsc{FiniteFlow}~\cite{Peraro:2019svx} can help to reduce the time needed to solve resulting linear systems. Of course, with the increasing tensor rank and number of legs, the traditional reduction becomes increasingly impractical, making it worth looking at alternative methods. However, as long as the complexity is not an issue and the obtained results are useful, there is nothing wrong with using the traditional reduction for the given calculation.

Unfortunately, the traditional reduction has the unpleasant property of featuring inverse Gram determinants made of external momenta.
At special kinematic points (\eg $p^2=0$ for a two-point function) such Gram determinants vanish, leading to the complete breakdown of the reduction procedure. In the past, this problem has been addressed \eg in the context of one-loop Passarino--Veltman reduction~\cite{Denner:2005nn,Patel:2015tea}, where one can avoid inverse Gram determinants by modifying the process of expressing  Passarino--Veltman coefficient functions in terms of scalar functions. An alternative solution where the reduction is set up using an auxiliary vector and the divergences from inverse Gram determinants are required to cancel in the final reduction coefficients has been presented in ref.~\cite{Feng:2022rfz}. Let us also remark that \textsc{AmpRed} can handle zero Gram determinant tensor integrals by reducing their Feynman parametric representation and using IBPs~\cite{Chen:2019fzm}.

In this work we would like to present yet another solution for this problem, that also makes use of IBPs and auxiliary vectors. Although the underlying idea certainly must be familiar to some practitioners, we are not aware of a literature source where it has been described in details. Judging from the amount of questions related zero Gram determinants we received from \textsc{FeynCalc} users in the past, practical knowledge of how to handle such cases does not appear to be widespread and so it is worth writing it down. The main idea here is to address the root cause for the vanishing Gram determinant, which comes from a linear dependence between external momenta present in the tensor integral. Let us elaborate on this approach by discussing three typical cases of zero Gram determinants

If two external momenta $p_1$ and $p_2$ are linearly dependent and not light-like, then they must satisfy
\begin{equation}
p_2^\mu = c p_1^\mu, \label{eq:tred-p1p2}
\end{equation}
where the constant $c$ can be determined from the process kinematics. This knowledge allows us to simplify the construction of the tensor basis by considering only $p_1$, but not $p_2$. The resulting tensor decomposition, being free of $p_2$ in the tensor coefficients, will lead to a solvable system of linear equations free of zero Gram determinants. Yet, there are some caveats. Upon doing the reduction we may encounter scalar products of the loop momenta $k_i$ and $p_2$ stemming from the integral denominators. Those quantities can be eliminated using Eq.\eqref{eq:tred-p1p2} since
\begin{equation}
k_i \cdot p_2  = c \, k_i \cdot p_1.
\end{equation}
Notice that this case can be completely handled using the traditional tensor reduction without using IBPs or auxiliary vectors. The only drawback is that the tensor structure of the result will only contain $p_1^\mu$, but not $p_2^\mu$.

Let us also consider the situation where there is only one vanishing momentum $p^2=0$. In this case there are no momenta to eliminate so that the method described above does not seem to be useful. The problem here is that one light-like vector turns out to be insufficient to construct a tensor basis. We can circumvent this issue by introducing an auxiliary vector $n$ and doing the tensor decomposition with respect to $p$ and $n$. For convenience, we may choose $n^2=0$ as long as $p \cdot n \neq 0$. The so-obtained results will depend on the inverse Gram determinant $(n \cdot p)^{-1}$ and also contain scalar integrals with $k_i \cdot n$ in the numerator. Naively, this might seem troublesome, since tensor decomposition of the original integral may not depend on an arbitrarily chosen vector $n$. However, the dependence on $n$ merely stems from the fact that the tensor decomposition has not yet been finished. While we succeeded in expressing our zero Gram determinant tensor integral in terms of scalar loop integrals, albeit at the cost of introducing $n$, the resulting scalar integral can still be reduced even further using IBP technology. To set up IBP reduction for integrals with $k_i \cdot n$ in the numerator, we need to augment the underlying integral families with auxiliary propagators containing $n$ so that the propagators form a basis. The obvious choice would be $1/(k_i \cdot n)$.
Upon running the reduction and inserting the results into our reduced tensor integral we see that all dependence on $n$ cancels as it should.

Another interesting example for a singular kinematic configuration would be $p_1^2 = p_1 \cdot p_2$ = 0 with $p_2^2 \neq 0$. In this case 
it is easy to see that $p_1$ vanishes identically, since
\begin{align}
    p_1^\mu = c p_2^\mu \Rightarrow  0 =  p_1 \cdot p_2 = c \underbrace{p_2^2}_{\neq 0} \Rightarrow c = 0.
\end{align}
Thus, setting $p_1$ to zero in the integral gets the tensor reduction done with any further issues.

In realistic calculations one may encounter a mixture of all three cases, where the Gram determinant vanishes because some of the external momenta are linearly dependent, while others might be light-like or identically zero. However, by resolving all linear dependencies, eliminating vanishing momenta and possibly introducing an auxiliary vector $n$ one can always construct a linearly independent tensor basis out of the original momenta. To streamline this procedure as much as possible,  \textsc{FeynCalc} 10.2 features a new routine called \texttt{FCLoopFindTensorBasis} which, given a set of four-momenta and their kinematics, can automatically resolve any linear dependencies between those vectors and suggest a new basis for tensor reduction.
The momenta in this new set are guaranteed to be linearly independent so that the corresponding tensor reduction will lead to a solvable system of linear equations. Furthermore, functions responsible for tensor reduction such as \texttt{TID}, \texttt{FCMultiLoopTID} and \texttt{FCLoopTensorReduce} have received a new option \texttt{TensorReductionBasisChange} which allows the user to enforce a tensor basis change according to the output of \texttt{FCLoopFindTensorBasis}.

This way the year-long problem of zero Gram determinants in calculations done with \textsc{FeynCalc} can be now handled in a semi-automatic fashion. The only price to pay is the appearance of asymmetric tensor bases (because some of the original momenta will be eliminated) and the need to do IBP reduction even at one-loop for some kinematic configurations. On the other hand, this method is conceptually clear and easy to implement in other codes, so that it may be a natural choice for all projects where the traditional tensor reduction is sufficient for the given calculation, but the process kinematics leads to zero Gram determinants.

\section{Extraction of UV-divergences and renormalization} \label{sec:renormaliz}

Conventional dimensional regularization does not distinguish between ultraviolet (UV) and infrared (IR) divergences, 
mapping both to poles in $1/\varepsilon$. In most practical applications this feature does not cause any issues, unless we 
are interested in calculating renormalization constants and need to extract UV-poles only.

For minimal subtraction ($\textrm{MS}$ and $\overline{\textrm{MS}}$) schemes, where the UV-counterterms
are known to be polynomial in momenta and masses, these problems can be circumvented using so-called
``infrared rearrangement''~\cite{Vladimirov:1979zm,Chetyrkin:1980pr} and $R^\ast$-operation~\cite{Chetyrkin:1984xa,Smirnov:1985yck}. Both techniques heavily rely on analyzing the involved Feynman diagrams and augmenting some of their lines with auxiliary masses and momenta in order to regulate IR-divergences. This makes their practical application in computer codes somewhat nontrivial, especially for nonexperts.

A more practical and easy-to-use algorithm was introduced in refs.~\cite{Misiak:1994zw,Chetyrkin:1997fm} and is commonly known under the name of ``tadpole expansion''. The main idea here is to exploit the exact relation
\begin{align}
    \frac{1}{(k-q)^2} = \frac{1}{k^2 - m^2} - \frac{q^2 - 2 k \cdot q + m^2}{k^2 - m^2} \frac{1}{(k-q)^2}, \label{eq:tadex}
\end{align}
with $m$ being an auxiliary mass introduced to regulate IR divergences in massless propagators. Notice that while the last term still contains the original massless denominator, the first term depends only on a massive denominator free of external momenta. In fact, \Eq\eqref{eq:tadex} can be iterated multiple times, by applying it to the massless denominator in the last term of the resulting equation. The number of times this recursion needs to be repeated is determined by the superficial degree of divergence $\omega(\Gamma)$ of the given Green function. Higher order terms free of UV-divergences (that still contain the external momentum $q$ in the denominator) are then discarded. For a 4-dimensional theory the formula for the superficial degree of divergence can be written as~\cite{Weinberg:1995mt}
\begin{align}
\omega(\Gamma) = 4 - \sum_f E_f (s_f + 1) - \sum N_i \Delta_i,
\end{align}
where $E_f$ is the number of external lines of $f$-type fields and $N_i$ denotes the number of $i-$type vertices, while $\Delta_i$ specifies the dimensionality of their coupling constants. Although $s_f$ can be interpreted as the spin of the field, in the case of QED and QCD we find $s_f=0$ for photons and gluons due to the gauge invariance. For example, for QCD we have
\begin{align}
    \omega_\textrm{QCD}(\Gamma) = 4 - (E_g + E_u) - \frac{3}{2} E_q,
\end{align}
with $g$, $u$ and $q$ being gluons, ghosts and quarks respectively. 

A diagram with $\omega(\Gamma) < 0$ is considered to be UV-convergent, provided that possible UV-subdivergences have been correctly subtracted. Subdivergences start to appear beyond one loop and need to be removed by including loop corrections to the UV-counter\-terms and extracting their UV-divergences (\eg using tadpole expansion).

In ref.~\cite{Chetyrkin:1997fm} it was observed that in order to simplify the calculation one may also neglect contributions proportional to $m^2$ in the numerators of  \Eq\eqref{eq:tadex} from the start. While these terms are necessary to ensure that our results do not depend on the auxiliary mass, they can be mimicked by including  additional counter-terms proportional to $m^2$ to the Lagrangian of the theory.  If the Lagrangian contains only dimension-4 operators, the counter-terms will essentially look like mass terms for massless vector bosons\footnote{An additional fermion ``mass term'' $m^2 \bar{\psi} \psi$ would be dimension 5.}, e.g. $m^2 G^a_{\mu \nu} G^{a \mu \nu}$ for a gluon. It is important to stress that these terms should be understood as a purely technical trick to correct the mistake one does by neglecting $m^2$-terms in \Eq\eqref{eq:tadex}. They do not modify the physics of the underlying theory (\eg by violating gauge invariance) and the renormalization constants computed this way are universally valid.

Unfortunately, even when omitting $m^2$-terms, tadpole expansion as described in ref.~\cite{Chetyrkin:1997fm} remains rather cumbersome to implement in practice. The main reason for this is that truncating \Eq\eqref{eq:tadex} at the desired order $\omega(\Gamma)$ breaks the invariance of loop diagrams under loop momentum shifts. In particular, as explained \eg in ref.~\cite{Lang:2020nnl}, to avoid inconsistencies one must ensure that the given diagram and the corresponding counter-terms with loop corrections have their momenta routed in the same way.

A more practical recipe for using this technology can be found \eg in ref.~\cite{Zoller:2014xoa}. Instead of using \Eq\eqref{eq:tadex} directly, one can simply add the auxiliary mass to every massless propagator and then expand the diagram in the external momentum and physical masses up to $\omega(\Gamma)$. This can be motivated by observing that iterating \Eq\eqref{eq:tadex} and setting all $m^2$-terms to zero, precisely corresponds to expanding $1/[(k-q)^2 - m^2]$ in $q$. Although this procedure can be further optimized by removing some unnecessary terms that arise during the expansion of massive propagators (\cf \eg refs.~\cite{Lang:2020nnl,Stockinger:2023ndm,vonManteuffel:2025swv,Weisswange:2025tab}), the main idea of introducing auxiliary masses, expanding in $q$ and physical masses and compensating for the missing $m^2$-terms using counter-terms remains unchanged.

Hence, this modified tadpole expansion can be regarded  as a convenient tool to renormalize theories with quadratic propagators\footnote{The method is not applicable to eikonal propagators~\cite{Chetyrkin:1997fm}.} in minimal subtraction\footnote{Since tadpole expansion changes finite parts of the integrals, one cannot use this approach for schemes where those pieces matter.} schemes. To the best of our knowledge, the only public tool that implements this technology is \textsc{MaRTIn}~\cite{Brod:2024zaz}, that was used \eg in renormalization calculations of  refs.~\cite{Aebischer:2023djt,Steudtner:2024teg,Steudtner:2025blh,Schroder:2025rka,Brod:2026yxw}. However, since this package was developed for a very particular type of calculations where the user ends up with vacuum topologies, it may not be widely known among the broader particle-theory community.

In order to make the tadpole expansion method more accessible to everyone, we implemented this approach in \textsc{FeynCalc} by including suitable routines and adding some illustrative examples. First of all, using \texttt{FCLoopGetFeynAmpDenominators} one can extract all propagator denominators present in the given amplitude and wrap them into given head in the original expression. This makes it easy to change the propagators as needed and then plug them back into the amplitude. The insertion of auxiliary masses is handled via \texttt{FCLoopAddAuxiliaryMass}. Finally, one can expand the amplitude in the given four-momenta with the aid of \texttt{FourSeries}.

Although in the current version one can only expand in external momenta and not in particle masses, we believe that for calculations manageable with \textsc{FeynCalc} (mostly up to 2, rarely 3 loops) this should not constitute a serious limitation. Once the IBP-reduction is finished, one can always perform an asymptotic expansion of the tadpole master integrals to arrive at even simpler single-scale masters.
Using this technology we successfully reproduced literature results for the renormalization constants of QCD and QED up to 2 loops.

\section{Detection of factorizing integrals} \label{sec:factints}

Factorizing integrals are products of lower-loop integrals that frequently appear in the outcome of IBP reductions. Once identified, they are generally much easier to evaluate than genuine $x$-loop integrals. However, detecting and separating them by hand can be tedious so that an automated approach is highly desirable here. For master integrals without numerators, a well-known criterion is based on the Symanzik  $U$-polynomial: if the  $U$-polynomial factorizes, then the corresponding integral factorizes as well. The remaining simpler task is then to determine which loop momenta belong to the individual subintegrals and to rewrite the original integral as a product of simpler master integrals.

To streamline this procedure in \textsc{FeynCalc}, particularly when working with lists of master integrals produced by, \eg \textsc{FIRE} or \textsc{KIRA}, we have introduced the function \texttt{FCLoop\-FactorizingQ}. For integrals without numerators, it returns \texttt{True} if the integral factorizes and \texttt{False} otherwise. The actual decomposition is then performed by \texttt{FCLoop\-Create\-Factorizing\-Rules} (for integrals in \texttt{GLI}-notation) and \texttt{FCLoopFactorizingSplit} (for integrals in \texttt{FAD}-notation).

As part of this implementation, we also introduced the auxiliary function \texttt{FCLoopToGLI}. Given a loop integral in \texttt{FAD}-notation, it converts it  into a \texttt{GLI} object together with the corresponding \texttt{FCTopology}. This functionality is also useful independently of factorization-related applications.

\section{Handling of incomplete and overdetermined topologies} \label{sec:topotreat}

While \texttt{FeynCalc} 10 introduced the long-awaited tools for topology identification and minimization, the treatment of certain special cases that commonly arise when constructing the final set of integral families remained less than satisfactory. In particular, difficulties occurred for topologies with incomplete or overdetermined sets of propagators, where the standard workflow of passing the output of \texttt{FCLoopFindTopologies} directly to \texttt{FCLoopFindTopologyMappings} was no longer sufficient.

To address these issues, we introduced the new routines \texttt{FCLoopRewrite\-Incomplete\-Topologies} and \texttt{FCLoop\-Rewrite\-Overdetermined\-Topologies}. Starting from the output of \texttt{FCLoop\-Find\-Topologies}, these functions automatically perform basis completion or partial-fraction decomposition, respectively, while consistently updating both the amplitude and the corresponding list of topologies.

As a result, the construction of a minimal set of topologies for a given amplitude can now be carried out in a much more automated and robust manner.

\section{Examples} \label{sec:examples}

When the first version of \textsc{FeynHelpers} was officially released in 2016, \textsc{FeynCalc} was mainly known as a package for tree- and one-loop level calculations. The add-on was partly born out of curiosity to see how far one could automatize fully analytic Feynman diagram evaluation with \textsc{FeynCalc} by directly interfacing it with the \textsc{Mathematica} version of \textsc{FIRE} and \textsc{Package-X}. Since then, \textsc{FeynCalc} became increasingly more popular in the multiloop community while the demand for reliable interfaces to IBP-reduction and sector-decomposition tools has continued to grow. Moreover, it has become clear that there are quite some phenomenologically relevant calculations which can be carried out efficiently by combining \textsc{FeynCalc} with \textsc{FeynHelpers}.

For this reason with the release of \textsc{FeynHelpers} 2.0 we decided to remove the somewhat artificial separation between examples that use \textsc{FeynCalc} alone and those that combine \textsc{FeynCalc} with \textsc{FeynHelpers}. To that aim we integrated \textsc{FeynHelpers} 2.0 into all suitable \textsc{FeynCalc} examples at one-loop and beyond to demonstrate how much simpler things can get when both codes (plus tools interfaced via \textsc{FeynHelpers}) are used together. As a consequence, most \textsc{FeynHelpers} examples can be now found in \textsc{FeynCalc}'s example directory under \texttt{FileNameJoin[\{\$FeynCalcDirectory, "Examples"\}]}.

In addition to that, we added a specific example showcasing how \textsc{FeynCalc} and \textsc{FeynHelpers} can be used for tackling practical multiloop research tasks. The calculation reproduces the two-loop amplitude for the decay $H \to gg$ via fermion loop presented in ref.~\cite{Anastasiou:2006hc}.  While the calculation proceeds via standard steps of simplifying Dirac and color algebra as well as reducing all loop integrals to masters, there are also several caveats. For example, to show that the reduced amplitude is gauge independent, it is necessary to uncover \emph{linear relations} between multiple master integrals. In this example this is done following the approach from ref.~\cite{Davies:2024kvt}. Furthermore, to compare the amplitude from the literature with our result one needs to convert between two different bases of master integrals, which is also not entirely trivial. The corresponding notebook is included as an auxiliary file with this paper.

Furthermore, as has already been mentioned in Sec.~\ref{sec:renormaliz} we showcase the renormalization of QED and QCD up to two loops in the minimal subtraction scheme using tadpole expansion.

\section{Summary} \label{sec:summary}

In this work we presented new versions of \textsc{FeynCalc} and \textsc{FeynHelpers} that allow to organize multiloop calculations
with \textsc{Mathematica} while seamlessly using state-of-the art codes for relevant evaluation steps. In particular, when employing \textsc{FeynHelpers}, users of \textsc{FeynCalc} can benefit from diagram generation with \textsc{QGRAF}, IBP-reduction with \textsc{FIRE} or \textsc{KIRA} and loop integral evaluation using \textsc{FIESTA} or \textsc{pySecDec}. The run cards for the corresponding programs can be created without leaving the current \textsc{Mathematica} notebook and their results can be directly loaded and translated into \textsc{FeynCalc} notation. This eliminates the error-prone manual conversion between different codes and makes the entire workflow as streamlined and user-friendly as possible.

To address the performance challenges that arise when using \textsc{FeynCalc} for algebraic simplifications of large expressions or for handling long lists of integrals and topologies, we have introduced a new mechanism that parallelizes selected functions across multiple CPU cores. This should allow for carrying out moderately complex calculations \ie with only a few scales and a limited number of diagrams, at the two- and even three-loop level using \textsc{FeynCalc} and \textsc{FeynHelpers} alone. While this may not sound impressive for multiloop practitioners, many particle theorists who wish to go beyond one loop but lack the required know-how may now benefit from the modern technology implemented in \textsc{FeynCalc} in an easy-to-use manner. 

As far as more realistic multiloop calculations are concerned, we are currently developing a new setup called \textsc{LoopScalla} that integrates \textsc{FORM} and \textsc{FeynCalc} with \textsc{FeynHelpers} into an automated framework for computing Feynman diagrams and expressing the results in terms of master integrals. A key feature of this code is its focus on distributing evaluations among multiple nodes on a computer cluster managed with \textsc{SLURM}. While we plan a stable in the near future, a preliminary working version of the code with minimal documentation is already publicly available \footnote{\url{https://github.com/FeynCalc/LoopScalla/}}.

Last but not least, we presented a pragmatic solution to the notorious problem of zero Gram determinants in multiloop tensor reduction.
The new FeynCalc function \texttt{FCLoop\-Find\-Tensor\-Basis} can automatically construct a set of linearly independent external momenta
required for a solvable tensor basis.  This approach is obviously also applicable to the one-loop Passarino--Veltman technique  and can be used to bypass the conventional way of first introducing the coefficient functions and then reducing them to scalar functions using formulas from the literature. Instead, one can employ the traditional tensor reduction implemented in \textsc{FeynCalc}'s \texttt{TID} function together with the IBP reduction to reduce any one-loop  tensor integral to scalars regardless of its kinematics. It would be highly beneficial to develop a public library of computer-readable analytic results for scalar one-loop integrals similar to what was implemented in \textsc{Package-X}. The demand for it still exists since the development of \textsc{Package-X} has officially been abandoned and the source code was never made open-source. Furthermore, \textsc{Package-X} never included higher orders in $\varepsilon$ that are important when substituting such results into multiloop amplitudes.

In the future, we would like to enhance \textsc{FeynHelpers} both in terms new interfaces and improvements of the existing ones. For example, it would be useful to add support for IBP reductions with \textsc{Blade} and numerical integral evaluation using \textsc{AMFlow}~\cite{Liu:2022chg}. Then, as far as \textsc{FIRE} and \textsc{KIRA} are concerned, \textsc{FeynHelpers} still cannot directly use these tools in the finite field mode.

All in all, we are convinced that even at the current development stage the combination of \textsc{FeynCalc} and \textsc{FeynHelpers} can be employed to conduct many interesting projects that require evaluation of Feynman diagrams at two or more loops. With the official release of the new versions, we look forward to receiving feedback from the community, which will help to further improve and extend these tools.

\section*{Acknowledgments}

The author would like to thank Guido Bell, Philipp Böer, Sudeepan Datta, Joshua Davies, Marvin Gerlach, Florian Herren, Dennis Horstmann, Tobias Huber, Aliaksei Kachanovich, Martin Lang, Fabian Lange, Vitaly Magerya, Alexander Smirnov, Matthias Steinhauser, Johann Usovitsch for valuable discussions and insightful comments that contributed to improving the multiloop capabilities \textsc{FeynCalc} and \textsc{FeynHelpers}. The author is also grateful to Tom Steudtner for his valuable insights into the intricacies of tadpole expansions.
VS's research was supported by the Deutsche Forschungsgemeinschaft (DFG, German Research Foundation) under grant 396021762 — TRR 257 “Particle Physics Phenomenology after the Higgs Discovery”. This paper has been assigned preprint numbers P3H-25-112 and  SI-HEP-2025-31.

\bibliographystyle{elsarticle-num}
\bibliography{paper.bib}

\end{document}